\documentclass[12pt]{iopart}

\usepackage{iopams}  

\usepackage{bm}% bold math
\usepackage[dvipdfmx]{graphicx} % add by Author (M.Nakata)

\newcommand{\mtr}[1]{\mathrm{#1} }

\newcommand{\odif}[2]{\ensuremath{\frac{d #1}{d #2 }}}

\makeatletter
\def\Hline{%
\noalign{\ifnum0=`}\fi\hrule \@height 1pt \futurelet
\reserved@a\@xhline}
\makeatother

\begin{document}

%\title{\LARGE{Landscape computations for the edge of chaos \\ in nonlinear dynamical systems}}
\title{Landscape computations for the edge of chaos in nonlinear dynamical systems}

\author{Motoki Nakata$^{1,2}$ and Masaaki Imaizumi$^{3,4}$}

\address{$^1$Faculty of Arts and Sciences, Komazawa University, 1-23-1 Komazawa, Setagaya, Tokyo 154-8525, Japan}
\address{$^2$RIKEN Interdisciplinary Theoretical and Mathematical Sciences Program (iTHEMS), 2-1 Hirosawa, Wako, Saitama 351–0198, Japan}
\address{$^3$Komaba Institute for Science, The University of Tokyo, 7-3-1 Hongo, Bunkyo, Tokyo 113-8654, Japan}
\address{$^4$RIKEN Center for Advanced Intelligence Project (AIP), 1-4-1 Nihonbashi, Chuo, Tokyo 103-0027, Japan}
%\ead{nakatamo@komazawa-u.ac.jp}

%%%%%%%%%%%%%%%%%%%%%%%%%%%%%%%%%%%%%%%%%%%%%%%%%%%%%%%%%%%%%%%%%%%%%%%%%%%
\begin{abstract}
We propose a stochastic sampling approach to identify stability boundaries in general dynamical systems. 
The global landscape of Lyapunov exponent in multi-dimensional parameter space provides transition boundaries 
for stable/unstable trajectories, i.e., the edge of chaos. 
Despite its usefulness, it is generally difficult to derive analytically. 
In this study, we reveal the transition boundaries by leveraging the Markov chain Monte Carlo algorithm coupled directly 
with the numerical integration of nonlinear differential/difference equation. 
It is demonstrated that a posteriori modeling for parameter subspace along the edge of chaos determines 
an inherent constrained dynamical system to flexibly activate or de-activate the chaotic trajectories.
\end{abstract}

%\maketitle

%\title[]{}

%%%%%%%%%%%%%%%%%%%%%%%%%%%%%%%%%%%%%%%%%%%%%%%%%%%%%%%%%%%%%%%%%%%%%%%%%%%
%
\section{Introduction}
Dynamical systems with a large number of degrees of freedom appearing in turbulence or many-body ensembles are 
often coupled each other. 
They involve several physical/control parameters that characterize the stability of trajectories through the nonlinearity. 
For instance, the Lorenz system to describe the low-dimensional but essential dynamics in turbulent convection\cite{lorenz} 
is widely recognized, where the 3 physical parameters determine the bifurcation of trajectory towards chaotic and/or stable periodic states. 
Another extreme examples are neural network systems.
In biological neural networks\cite{bioneuron0,bioneuron1,bioneuron2}, 
the nonlinear neuronal dynamics of firing or non-firing 
is described by strongly coupled differential or difference equations with several physical parameters for each. 
In artificial neural networks\cite{artneuron1,artneuron2}, linearly weighted superposition of input data with a nonlinear activation function
enables the high-performance learning by optimizing the network parameters such that minimize the training and validation loss. 
In either case, the individual parameter dependence of the trajectories tends to be complicated. 
Identifying the global structures of a critical quantity in multi-dimensional parameter space 
is fruitful to understand or extract the essential characteristics in a wide class of nonlinear dynamical systems. 
Indeed the two-dimensional landscape of the free energy functional has substantially contributed to the studies of molecular dynamics 
for protein folding\cite{protein}. 
Despite its usefulness, it is generally difficult to derive analytic expressions for the landscape of critical quantities 
from the governing equation. 
In this paper we propose a coupling computation to identify the global landscape of a general functional 
associated with the nonlinear dynamical system of interest. 
%
%The numerical integration of the dynamical equation is directly connected to the stochastic sampling in the multi-dimensional parameter space. 
%
Then we consider a one-dimensional chaotic neuron model as a representative, 
and revealed the Lyapunov stability landscape to characterize stable periodic or unstable chaotic trajectories, 
by leveraging the Replica Exchange Markov chain Monte Carlo (MCMC) sampling\cite{mcmc,rxmc}. 
It is also demonstrated that a posteriori modeling for the edge of chaos induces 
an inherent constrained dynamical system to activate the chaotic trajectories.
\section{Direct coupling of dynamical equation integration and stochastic sampling}
In this section we present a computational methodology to find the global landscape associated with the dynamical equation of interests. 
Numerical integrations of the dynamical equation are directly coupled with numerical explorations of the associated functional 
in terms of the stochastic sampling. 
\subsection{Formal framework of coupling computation}
Here we formally describe a nonlinear dynamical equation for a field $f(t,\bm{x})$ as follows: 
%
% Eq.1
\begin{equation} 
\partial_{t} f(t,\bm{x}) = \mathcal{N}(f,t,\bm{x}\,;\bm{\theta}), 
\end{equation}
%%%
where $\bm{x}$ and $t$ are, for instance, the $n+1$ dimensional spatio-temporal variables, 
and $\mathcal{N}$ is, in general, nonlinear function of $f$.
Control parameters to characterize the dynamical properties in the solution are given by 
$\bm{\theta} = (\theta_1,\theta_2,\cdots,\theta_m)\in \mathbb{R}^m$. 
In addition to the solution $f(t,\bm{x})$ under the appropriate initial and boundary conditions, 
we are often interested in the functional of $f$, i.e., $\mathcal{L}[f(\cdot);\bm{\theta}]$ 
which is a multi-dimensional function of $\bm{\theta}$. 
An integral type functional such as $\mathcal{L}[f(\cdot);\bm{\theta}] := \int dt d\bm{x}|f(t,\bm{x})|^2$ 
often appears in physics (e.g., the energy and entropy), and the local maxima and/or minima of $\mathcal{L}$ in the $\bm{\theta}$ space 
are of particular importance.     
It is however quite hard to derive analytically the functional form of $\mathcal{L}$ 
because of the nonlinearity of $\mathcal{N}$. 
Mathematical optimization algorithms such as the gradient descent methods are then utilized 
to calculate the extremum of $\mathcal{L}$ and its position in $\bm{\theta}$. 
Physically (and even mathematically) fruitful information is not only a single extremum itself, 
but also the distributions of extrema with their neighboring topography, 
because it contains the information of stability and robustness for the realizable solutions. 
To this end, we construct a coupled computation to identify the global landscape of $\mathcal{L}[f(\cdot);\bm{\theta}]$ 
by connecting the dynamical equation integration with the Markov chain Monte Carlo (MCMC) sampling\cite{mcmc}, 
as will be shown below.
\subsection{Nonlinear dynamical system as a representative}
As a representative dynamical system with multiple parameters, here we consider a one-dimensional equation 
to describe the firing activity of a single biological neuron\cite{artneuron1, artneuron2}, i.e., 
%
% Eq.2, Eq.3
\begin{eqnarray}
\xi(t+1)  =   k\xi(t) - \alpha U[\xi(t)] + a, \\
x(t+1) = U[\xi(t+1)],
\end{eqnarray}
%%%
%
where the activation function is given by $U(z) = [1+\exp(-z/\epsilon)]^{-1}$. 
% 
% Fig.1
\begin{figure}[b]
\centering
\includegraphics[scale=0.5]{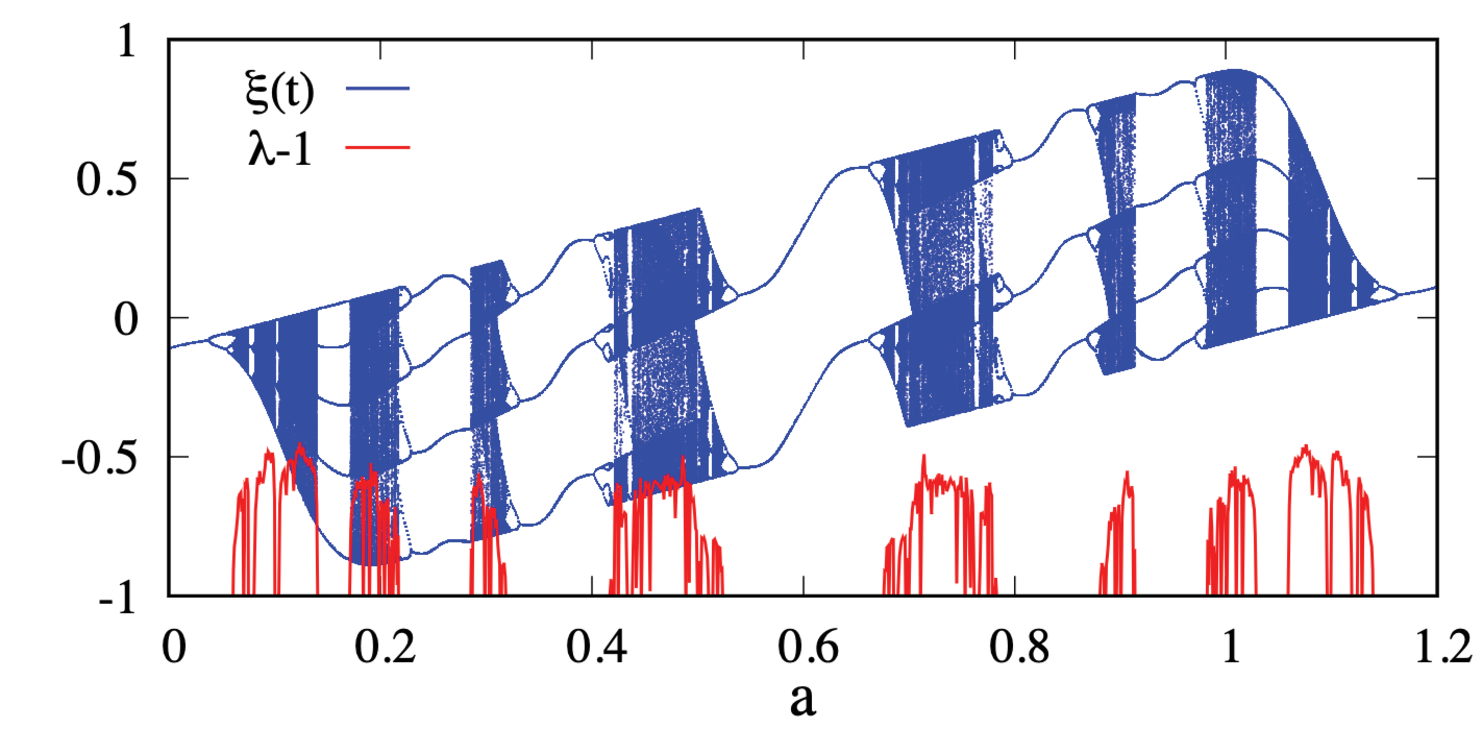}
\caption{
Bifurcation map with respect to the parameter $a$ in the chaotic neuron model. 
The asymptotic time evolutions of the internal state variable $\xi(t)$ are over-plotted at each value of $a$, 
where the other 3 parameters are fixed to $k \! = \! 0.85$, $\alpha \! = \! 1.2$, $\epsilon \! = \! 0.026$. 
The distribution of Lyapunov exponent $\lambda$ shifted with $-1$ is also displayed. 
}
\end{figure}
The internal state of neuron at the discrete time $t \! \in \! \mathbb{Z}^{+}$ is denoted by $\xi(t)$, 
and the output $x(t+1)$ takes either 1 (firing) or 0 (non-firing) value. 
The equations (2) and (3) are called the chaotic neuron model 
because they involve the chaotic trajectories in $\xi(t)$ depending on the 4 parameters $(a,k,\alpha,\epsilon)$, 
where the Lyapunov exponent is given by   
%
% Eq.4
\begin{equation} 
\lambda := \lim_{N\to \infty} \frac{1}{N}\sum_{t=1}^{N}\ln \left| \odif{\xi(t+1)}{\xi(t)} \right|. 
\end{equation}
%%%
%
%
% Fig.2
\begin{figure}[b]
\centering
\includegraphics[scale=0.5]{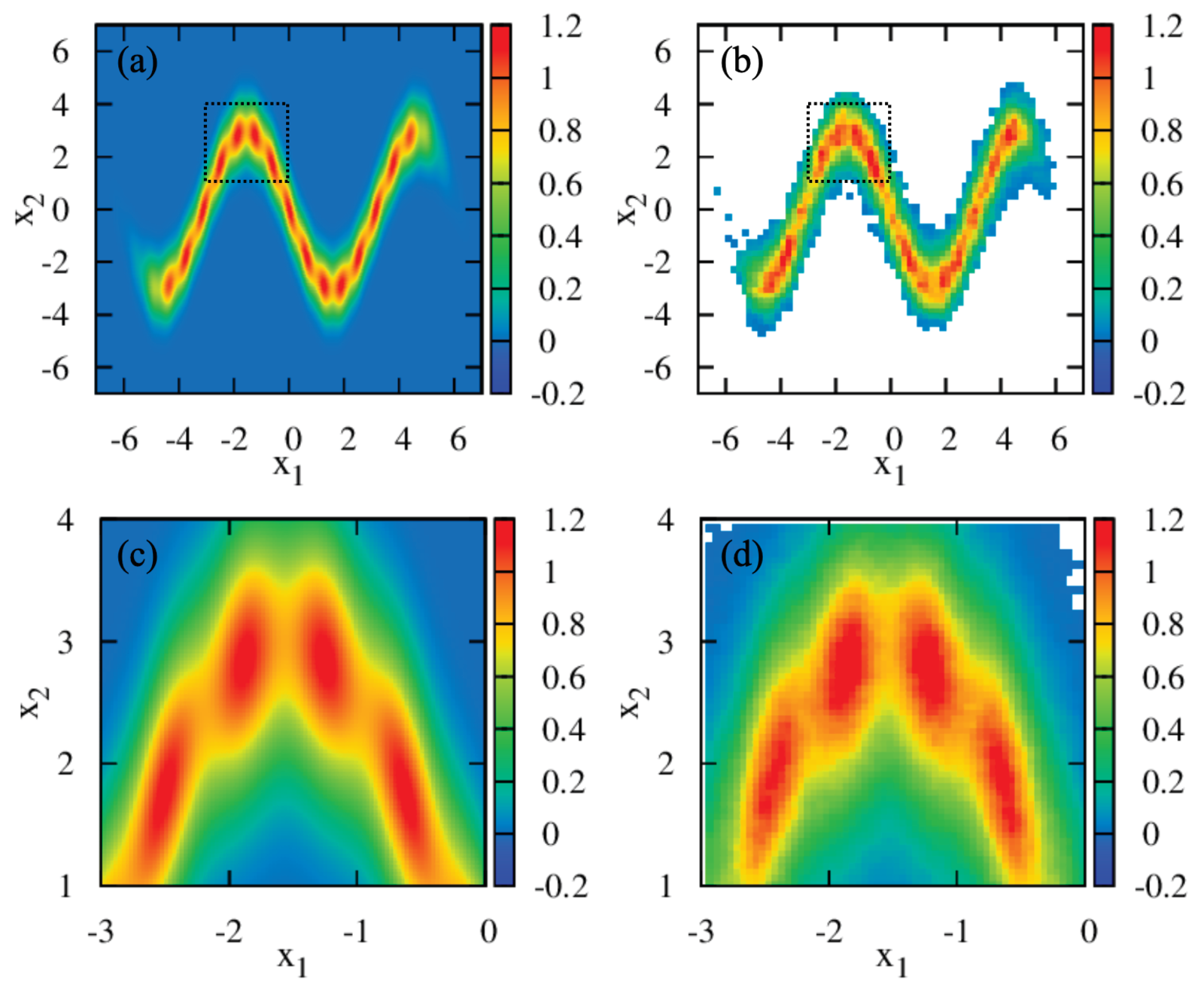}
\caption{
Comparisons of the functional form $T(x_1,x_2)$ in Eq. (5) which is numerically sampled by using [(a),(c)]grid search 
with $1000^2$ points and [(b),(d)]Replica Exchange MCMC (RXMC) algorithms with 50000 steps and 5 replicas. 
The results for the narrower region (indicated by the dashed square) are also shown in (c) and (d). 
}
\end{figure}
A typical bifurcation map with respect to the parameter $a$ is shown in Fig. 1, where the other 3 parameters are fixed. 
One can recognize the chaotic or unstable trajectories for $\xi(t)$ in the sub-region with $\lambda > 0$.
The Lyapunov exponent $\lambda$ indicates complicated structures even in such a simple dynamical system, 
and it is often difficult to identify the distribution of local extrema and zero boundaries 
in terms of optimization algorithms. 
It is thus worthwhile to connect directly with the stochastic sampling algorithm. 
\subsection{Replica Exchange MCMC sampling to be coupled}
We consider a stochastic sampling based on the Markov chain Monte Carlo (MCMC) algorithm 
as a numerical global search for the landscape of $\mathcal{L}[f(\cdot);\bm{\theta}]$, 
which is determined by the functional of the trajectory $f$ after integrating the dynamical equation. 
The Lyapunov stability landscape, i.e., $\mathcal{L}[\xi(t); (a, k, \alpha, \epsilon)]=\lambda$, will be discussed in the next section, 
and here we briefly summarize the Replica Exchange MCMC (abbreviated by RXMC) algorithms used in this study. 
Let $T(x_1,x_2)$ be a two-dimensional trial function with multi-modal and sparse characteristics: 
%
% Eq.5
\begin{eqnarray}
T(x_1,x_2) := & \ \frac{1}{2}\left [ 1 \! + l_1\cos(l_2x_1) \right ] \exp \! \left [-(l_3\sin x_1 \! + x_2)^2 \right ] \nonumber \\ 
 & \times \left [ \tanh(l_4 x_1 \! + l_5) - \tanh(l_4 x_1 \! - l_5) \right ], 
\end{eqnarray}
%%%
%
where $(l_1,l_2,l_3,l_4,l_5) = (0.15, 10, 3, 2, 10)$, and the contour map of $T(x_1,x_2)$ is shown in Fig. 2(a).  
MCMC algorithm can sample the multi-modal function by utilizing the Markov chain 
which is a stochastic process of discrete random variables $X_t$ (corresponding to $\bm{\theta}$) with the Markov property 
of $P[X_{t+1}|X_t,X_{t-1},X_{t-2},\cdots,X_1] = P[X_{t+1}|X_{t}]$. 
The Metropolis algorithm is applied to determine the acceptance probability 
for each Monte Carlo step, i.e., $R_{\mtr{accept}} = \min\{1,P[X_{t+1}]/P[X_t]\}$, 
then the histogram and expected values are obtained for $P$ as the stationary probability distribution to be sampled. 
In RXMC the additional parameter $\beta_j$ called the temperature parameter is introduced, 
and the family of stationary probability distribution is given by 
%
% Eq.5
\begin{equation} 
P(\bm{\theta}|\beta_j) = \exp \left [ \beta_j \mathcal{L}[f(\cdot);\bm{\theta}]  \right ] /\, Z(\beta_j), 
\end{equation}
%%%
%
where the integer $j$ stands for the index of the replicas, 
and $Z(\beta_j)$ is the normalization factor for each replica of $\beta_j$ such that $\int d\bm{\theta} P(\bm{\theta}|\beta_j) = 1$. 
Two random variables in different replicas are mutually exchanged at the period associated with $\beta_j$. 
For more details of RXMC algorithm, see for instance Ref. \cite{rxmc}. 
The sampling results for the trial function $T(x_1, x_2)$ in terms of the uniform grid search and RXMC algorithms are compared  
in Fig. 2. 
One finds that the RXMC case efficiently sampled the trial function only around their peak values.
Besides, the fine-scale multi-modal peak structures are also recovered when the sampling domain is bounded to be narrower. 

\section{Landscape computation}
Here we demonstrate the effectiveness of the direct coupling computation described above, 
where the numerical integration of the chaotic neuron model in Eqs. (2) and (3) is coupled with the RXMC sampling algorithm. 
Then the global Lyapunov stability landscape of $\mathcal{L}[f(\cdot);\bm{\theta}] = \lambda(a,k,\alpha,\epsilon)$ 
in the four-dimensional parameter space is calculated. 
We set the Monte Carlo steps of 100000 with the 10 replicas for the present coupled computation. 
At each replica, the numerical sampling with the integration of $\xi(t)$ for each point in $(a,k,\alpha,\epsilon)$ 
is carried out in parallel computations, and the exchange probability is evaluated in the period of 50 Monte Carlo steps. 
The three-dimensional landscape of the non-negative part in Lyapunov exponent $\lambda \geqslant 0$ is shown in Fig. 3(a), 
where $\alpha = 1.2$ is fixed for visualization. 
For comparison, the result in the case with $200^3$ grid search algorithm is displayed in Fig. 3(b). 
One finds that the present coupling computation with RXMC well captures the global and local structures in $\lambda$, 
as well as the edge of chaos given by $\lambda = 0$. 
Note that the uniform grid search redundantly scans the whole rectangular domains in $(a,k,\epsilon)$ 
whereas the coupled computation with RXMC scans only the limited domains with $\lambda \geqslant 0$ (colored domain in the figure). 
We also emphasize that the present method is effective even in the cases with non-differentiable functions, 
which can not be searched by the gradient-based optimization algorithms.  
%
%
%
% Fig.3 
\begin{figure}[h]
\centering
\includegraphics[scale=0.55]{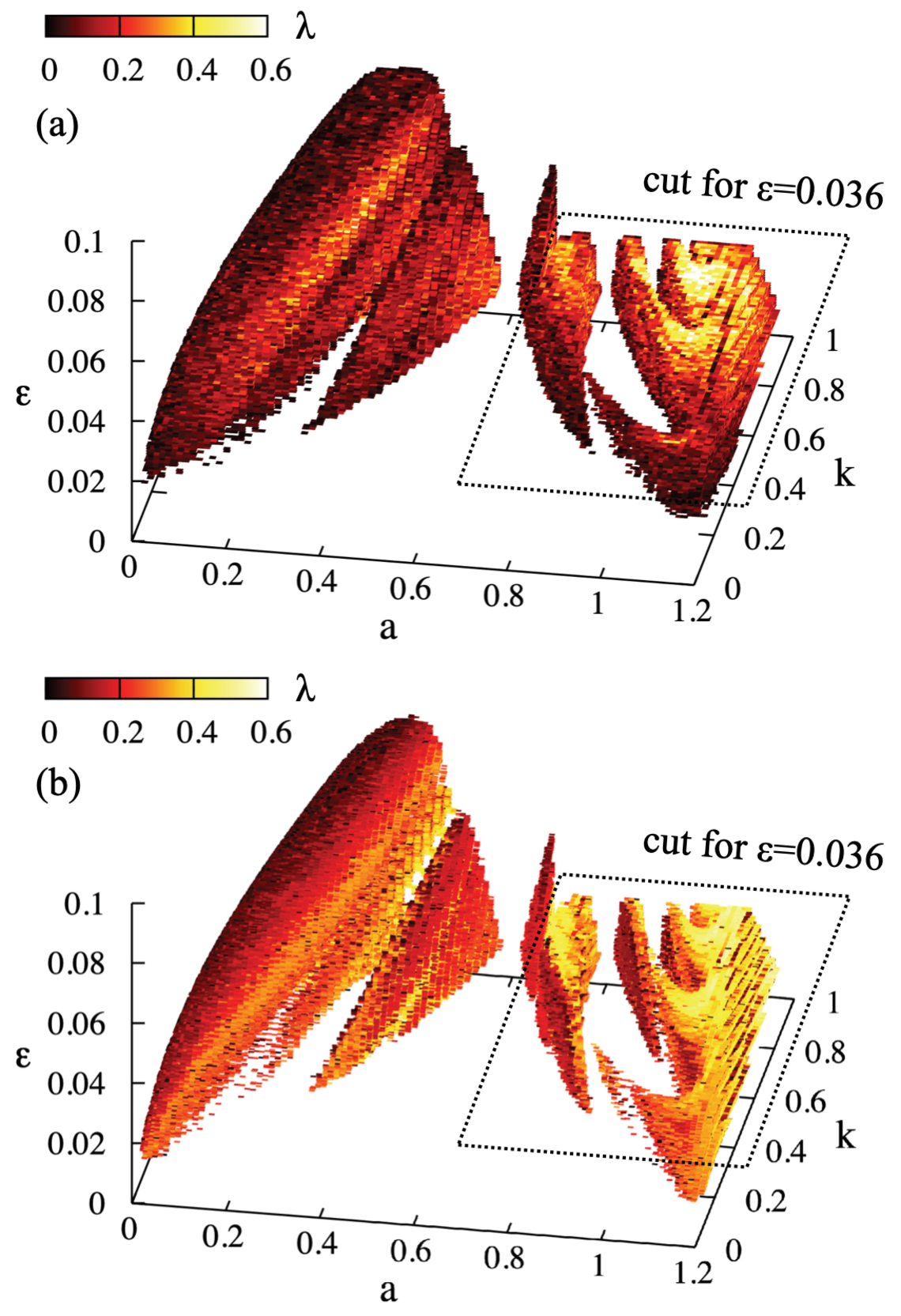}
\caption{
Global landscape of the non-negative Lyapunov exponent $\lambda(a,k,\epsilon,\alpha) \! \geqslant \! 0$ 
obtained by (a)grid search and (b)RXMC algorithms, where $ \alpha \! = \! 1.2$ is fixed for the visualization. 
The values in the sub-region of $a \! > \! 0.7$ and $\epsilon \! > \! 0.036$ are not displayed to make the internal structures visible. 
}
\end{figure}
\section{A posteriori modeling for the edge of chaos}
Once the Lyapunov stability landscape for the nonlinear dynamical system of interest is obtained, 
one can construct an inherent constrained dynamical system to flexibly activate or de-activate the chaotic trajectories.
In this section, we briefly demonstrate it by carrying out a posteriori modeling for the edge of chaos in the landscape. 
The two-dimensional subspace of $(a,k)$ at $\epsilon = 0.036$ is considered here for simplicity, 
and the contour map of $\lambda$ is shown in Fig. 4. 
Then one can construct an arbitrary approximate curve $g$ along the edge of chaos shown by the black solid line in the figure, 
where the analytic expression is given by 
%
% Eq.6
\begin{equation} 
k = g^{-1}(a) := c_1(a-c_2) + \sqrt{c_1(a-c_2)} + c_3, 
\end{equation}
%%%
%
with the parameters of $c_1 \! = \! -4.0$, $c_2 \! = \! 1.137$, $c_3 \! = \! 0.3$. 
While we determined the functional form of $g^{-1}$ in a heuristic manner from the landscape geometry, 
other systematic methods such as symbolic regression\cite{symbol} can be also utilized. 
Since the Lyapunov exponent $\lambda$ is non-negative at the most points along the curve $g$, 
such a one-dimensional curve in the two-dimensional $(a,k)$ space provides 
a non-trivial parameter constraint such that the trajectory of $\xi(t)$ becomes unstable or chaotic. 

By applying the parameter constraint $a=g(k)$ to the dynamical equation (2), 
one obtains an inherent constrained dynamical system, that is formally given as: 
%
% Eq.7
\begin{equation}
\xi(t+1)  =   k\xi(t) - \alpha U[\xi(t)] + g(k), \\
\end{equation}
%%%
where the parameter $a$ is now eliminated from the dynamical equation. 
Figures 5(a) and 5(b) compare the bifurcation maps between the standard case with Eq. (2) and the constrained case with Eq. (8), respectively.   
One can see that, as is expected from the constraint in Eq. (7), the chaotic trajectory in Eq. (8) 
is activated for almost all region in the parameter $k$. 
It is noteworthy that one can also de-activate flexibly the chaotic trajectory to the stable periodic behaviour  
if the constraint is slightly modified to deviate from the edge of chaos. 
%
%
% Fig.4
\begin{figure}[t]
\centering
\includegraphics[scale=0.55]{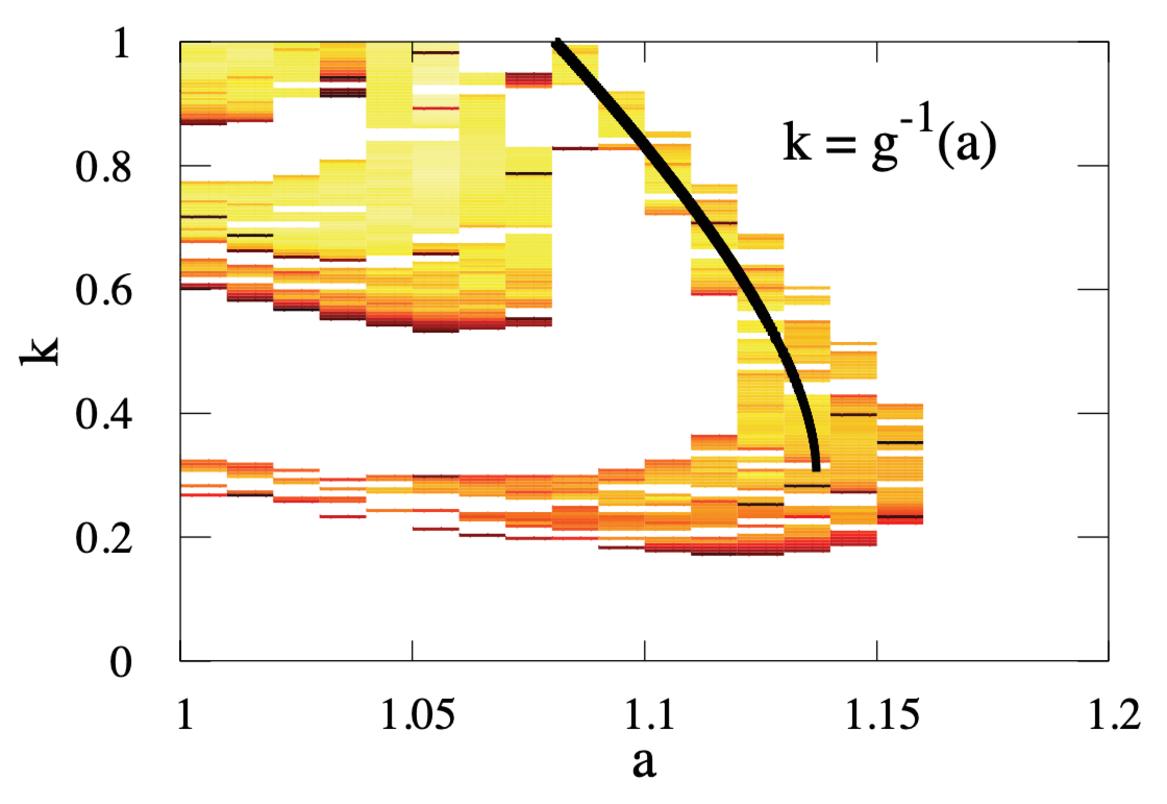}
\caption{
Two-dimensional local landscape of the non-negative Lyapunov exponent $\lambda \! \geqslant \! 0$ 
on the cross section at $\epsilon \! = \! 0.036$ (shown by dashed squares in Fig. 3). 
A one-dimensional curve determined by the model function $g$ along the outer edge of the landscape is also plotted by the black solid line. 
}
\end{figure}
%
% Fig.5
\begin{figure}[t]
\centering
\includegraphics[scale=0.45]{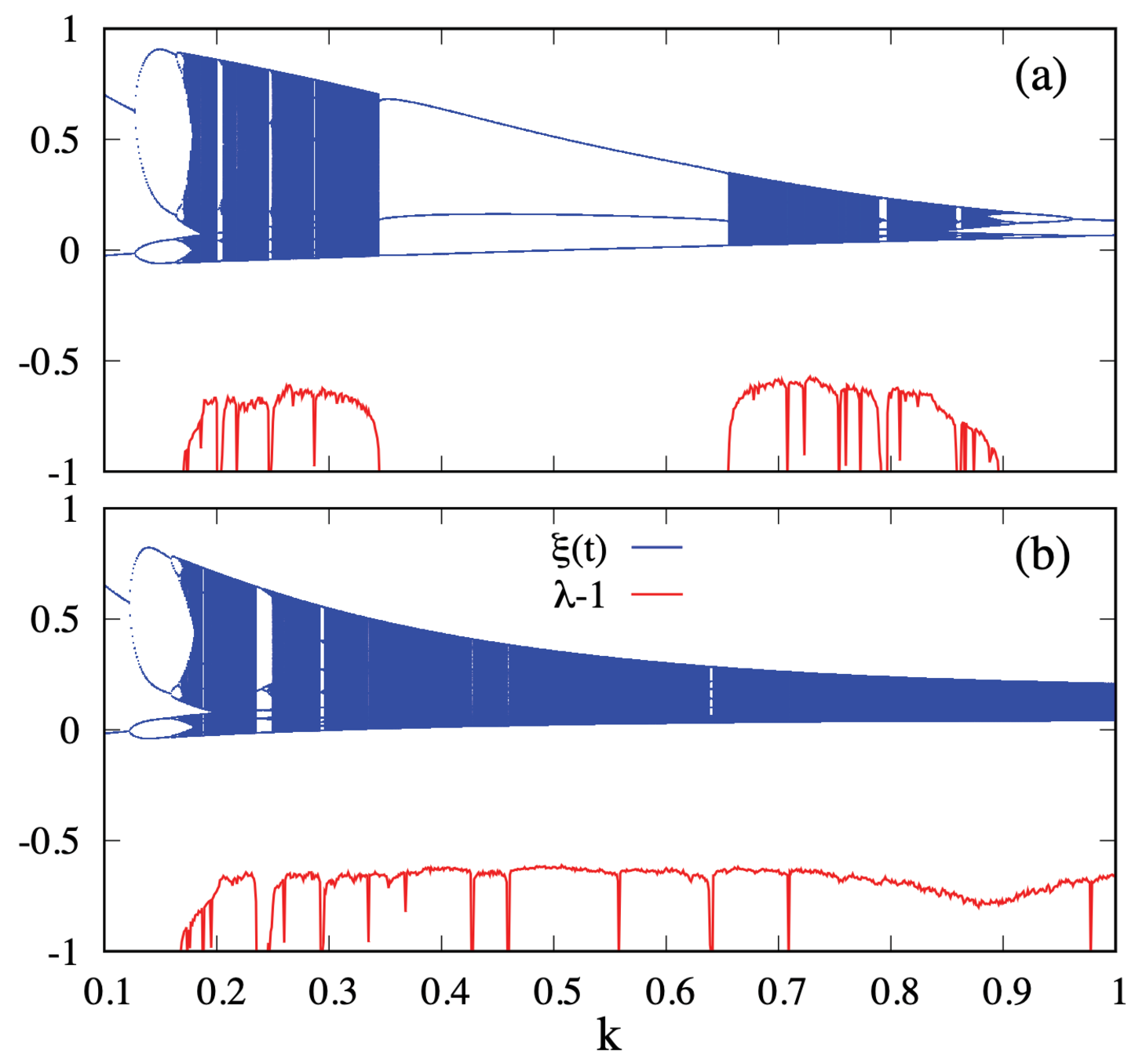}
\caption{
Comparison of the bifurcation maps between (a)the original chaotic neuron model [Eqs. (2) and (3)] 
with the constant parameter of $a\! = \! 1.1$ and (b)the extended model [Eqs. (8) and (3)] 
with the parameter constraint of $a \! =\! g(k)$ in Eq. (7). 
The other parameters are fixed to $\alpha \! = \! 1.2$, $\epsilon \! = \! 0.036$. 
The distributions of Lyapunov exponent $\lambda$ shifted with $-1$ are also plotted. 
}
\end{figure}
\section{Summary}
A direct coupling computation to connect the dynamical equation integration to Markov chain Monte Carlo sampling 
is proposed in this paper.  
As a representative application, the multi-dimensional Lyapunov stability landscape associated with the chaotic neuron model 
is identified by using the Replica Exchange MCMC algorithm, 
where the global distribution of the Lyapunov exponent is efficiently scanned.    
We also demonstrate that a posteriori modeling for the parameter subspace along the edge of chaos on the landscape 
can construct an inherent constrained dynamical system to activate or de-activate the chaotic trajectories.
The fundamental methodology proposed here is also applicable to broader dynamical systems, 
even in the cases with non-differentiable functionals.
In this paper, we focused on the one-dimensional dynamical system in order to express the conceptual perspective. 
As for the higher dimensional cases, e.g., pattern formations in fluid and plasma turbulence\cite{plasma1, plasma2}, 
coupled nonlinear oscillators, neural networks, will be addressed in future works. 
\section*{Acknowledgments}
The author(M.N.) thanks Dr. Yoshinobu Kawahara for fruitful discussions on this work.  
This work is supported in part by JST, PRESTO Grant Number JPMJPR21O7, 
and in part by the NIFS collaborative Research programs (NIFS23KIST039).
Numerical computations were performed by Plasma Simulator at NIFS. 
%
%
%\vspace{-0.25cm}
%%%%%%%%%%%%%%%%%%%%%%%%%%%%%%%%%%%%%%%%%%%%%%%%%%%%%%%%%%%%%%%%%%%%%%%%%%%%%%%%%%%%%%%%%
%%%%%%

%
%

\section*{References}
\vspace{-0.2cm}
\begin{thebibliography}{11}
\bibitem{lorenz}
E. N. Lorenz, 
Deterministic Nonperiodic Flow, 
J. Atomos. Sci., {\bf 20} (1963), 130--141.
%
\bibitem{bioneuron0}
R. FitzHugh, 
Impulses and physiological states in theoretical models of nerve membrane, 
Biophysical J., {\bf 1} (1961), 445--466.
%
\bibitem{bioneuron1}
J. Nagumo and S. Sato, 
On a Response Characteristic of a Mathematical Neuron Model, 
Kybernetik, {\bf 10} (1972), 155--164.
%
\bibitem{bioneuron2}
K. Aihara, T. Takabe and M. Toyoda, 
Chaotic neural networks, 
Phys. Lett. A, {\bf 144} (1990), 333--340.
%
\bibitem{artneuron1}
A. Krizhevsk, I. Sutskever and G. E. Hinton, 
ImageNet Classification with Deep Convolutional Neural Networks, 
Communications of the ACM., {\bf 60} (1990), 84--90.
%
\bibitem{artneuron2}
M. Imaizumi and K. Fukumizu, 
Deep Neural Networks Learn Non-Smooth Functions Effectively, 
Proc. Mach. Learn. Res., {\bf 89} (2019), 869--878.
%
\bibitem{protein}
J. N. Onuchic, Z. Luthey-Schulten, P. G. Wolynes, 
Theory of protein folding: the energy landscape perspective, 
Ann. Rev. Phys. Chem., {\bf 48} (1997), 545--600.
%
\bibitem{mcmc}
W. K. Hastings, 
Monte Carlo Sampling Methods Using Markov Chains and Their Applications, 
Biometrika, {\bf 57} (1970), 97--109.
%
\bibitem{rxmc}
K. Hukushima and K. Nemoto, 
Exchange Monte Carlo method and application to spin glass simulations, 
J. Phys. Soc. Jpn., {\bf 65} (1996), 1604--1608.
%
\bibitem{symbol}
M. Schmidt and H. Lipson, 
Distilling free-form natural laws from experimental data, 
Science, {\bf 324} (2009), 81--85.
%
\bibitem{plasma1}
M. Nakata, T. -H. Watanabe and H. Sugama, 
Nonlinear entropy transfer via zonal flows in gyrokinetic plasma turbulence, 
Phys. Plasmas, {\bf 19} (2012), 022303-1--14.
%
\bibitem{plasma2}
M. Nakata, T. -H. Watanabe, H. Sugama and W. Horton
Effects of parallel dynamics on vortex structures in electron temperature gradient driven turbulence, 
Phys. Plasmas, {\bf 18} (2011), 012303-1--11.
%
\end{thebibliography}
\end{document}